\documentclass[aps,superscriptaddress,twocolumn,showpacs]{revtex4-1}
\usepackage{amsmath}
\usepackage{color}
\usepackage{pict2e}
\usepackage{graphicx}
\usepackage{hyperref}

\begin{document}

\title{Spinon walk in quantum spin ice}
\author{Yuan Wan}
\affiliation{Perimeter Institute for Theoretical Physics, Waterloo, Ontario N2L 2Y5 Canada}
\author{Juan Carrasquilla}
\affiliation{Perimeter Institute for Theoretical Physics, Waterloo, Ontario N2L 2Y5 Canada}
\author{Roger G. Melko}
\affiliation{Perimeter Institute for Theoretical Physics, Waterloo, Ontario N2L 2Y5 Canada}
\affiliation{Department of Physics and Astronomy, University of Waterloo, Ontario, N2L 3G1, Canada}

\date{\today}

\begin{abstract}

We study a minimal model for the dynamics of spinons in quantum spin ice. The model captures the essential strong coupling between the spinon and the disordered background spins. We demonstrate that the spinon motion can be mapped to a random walk with an entropy-induced memory in imaginary time. Our numerical simulation of the spinon walk indicates that the spinon propagates as a massive quasiparticle at low energy despite its strong coupling to the spin background at the microscopic energy scale. We discuss the experimental implications of our findings.

\end{abstract}


\maketitle

Fractionalization is the phenomenon whereby elementary excitations are produced by breaking apart the constituent degrees of freedom of a system~\cite{Rajaraman2002}. It is manifest in experiments through the dynamic response of the fractional excitations. For example, perturbing a one-dimensional quantum Ising ferromagnet excites a pair of domain walls instead of a magnon. This results in a distinctive continuum as opposed to sharp resonance in neutron scattering spectra~\cite{Coldea2010}. In classical spin ice, the spins carry magnetic dipole moments~\cite{Bramwell2001}. The excitation is a mobile magnetic charge known as monopole, which can be regarded as a half of the dipole moment~\cite{Ryzhkin2005,Castelnovo2008,Castelnovo2012}. The diffusion of these monopoles gives rise to the unique magnetic relaxation behaviour of classical spin ice~\cite{Ryzhkin2005,Jaubert2009,Jaubert2011,Bovo2013}. Although crucial to the interpretation of experimental results on fractionalization, the dynamics of fractional excitations in two and three dimensions remains poorly understood. The microscopic models describing their dynamics are often strongly coupled. Thus, except for a small set of solvable models~\cite{Kitaev2006,Knolle2014}, the theoretical and numerical tools to tackle such problems are largely lacking.

Quantum spin ice (QSI) is a novel family of spin ice magnets with substantial quantum fluctuations~\cite{Gardner1999,Hermele2004,Molavian2007,Curnoe2007,Banerjee2008,Ross2011,Onoda2011,Savary2012,Lee2012,Applegate2012,Shannon2012,Benton2012,Dun2013,Gingras2014,Hallas2014,Hao2014,Huang2014,Henry2014,Kato2015,Lv2015}. The fractional excitations in QSI are spinons, which are the quantum analog of monopoles in classical spin ice. As opposed to a monopole, whose diffusion is the result of thermal fluctuations, the spinon propagates in QSI via quantum tunnelling. Central to the understanding of recent experiments probing the dynamics of QSI materials~\cite{Pan2016,Tokiwa2016}, the spinon dynamics has long been recognized as a challenging strong-coupling problem~\cite{Chen1974}. The spinon moves in a disordered spin background. The orientation of background spins controls the spinon motion, whereas the spinon motion in turn alters the spin background. One may naturally ask what a suitable framework for understanding the spinon dynamics is in QSI, and furthermore, whether there is a simple picture for the spinon propagation. In this Letter, we answer these questions by studying a minimal model that captures the essential features of single-spinon dynamics in QSI~\cite{Chen1974,Petrova2015}. We demonstrate that the spinon motion can be understood as a quantum walk with an entropy-induced memory. Equipped with this framework, we find that the spinon dynamics exhibits a remarkable renormalization phenomenon where the spinon propagates as a massive, nearly free quantum particle at low energy despite its strong coupling to a disordered spin background at the lattice scale.

\begin{figure}
\includegraphics[width=\columnwidth]{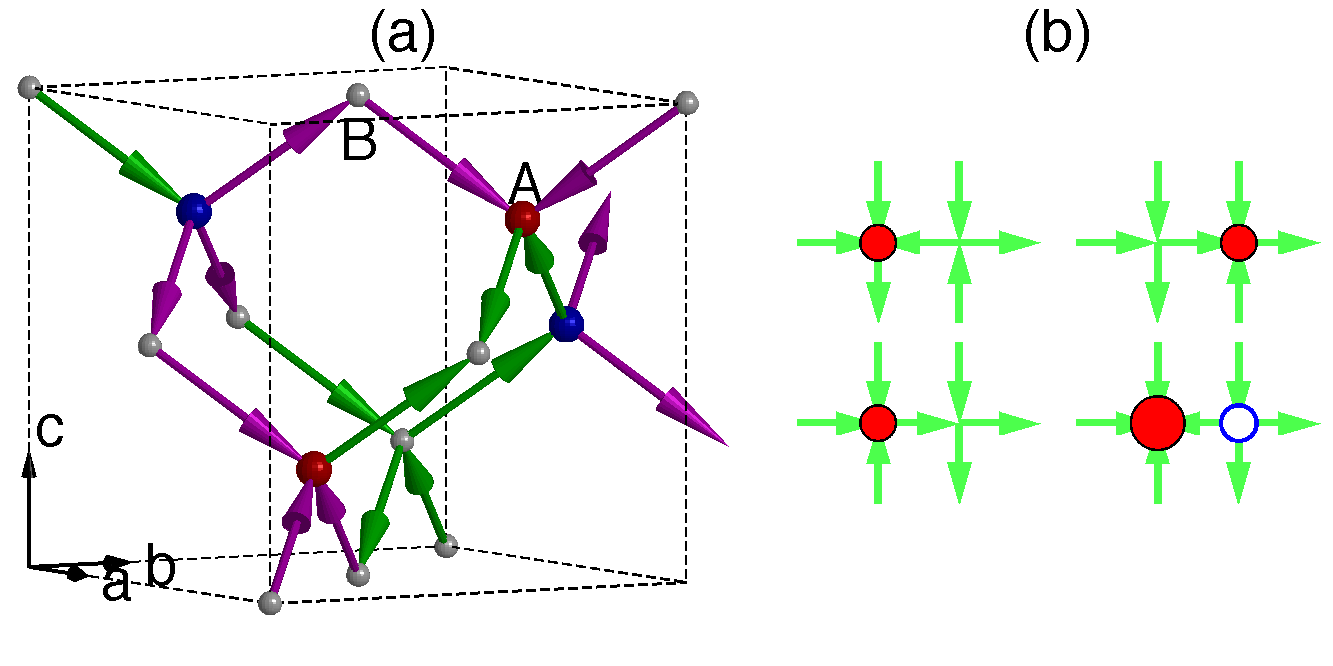}
\caption{(Colour online) (a) Spin ice model. Spins are located on the links of the diamond lattice. $Q=1$ and $-1$ spinons are shown as larger red and blue balls, respectively. The length of an edge of the unit cell (box) is scaled to 4.  Flippable spins are coloured in magenta. (b) Top: flipping the spin pointing toward the $Q=1$ spinon (red ball) brings it to the neighbouring site. Bottom: flipping the spin pointing away from the $Q=1$ spinon (red circle) creates a $Q=2$ spinon (large red circle) and a $Q=-1$ spinon (blue open circle).}
\label{pyro}
\end{figure}

QSI is best understood near the classical limit. We therefore start with the classical spin ice model~\cite{Bramwell2001}. To this end, we partition a diamond lattice into A and B sublattices. $S=1/2$ spins reside on the links. $S^z_{ij}\equiv S^z_{ji}=1/2\,(-1/2)$ if the spin on link $ij$ points from the A (B) site to the B (A) site. The classical model Hamiltonian is given by $H_\mathrm{CSI} = J/2\sum_{i}Q^2_i$, where the summation is over all lattice sites. $J>0$ is the exchange energy. $Q_i = \eta_i \sum_{j}S^z_{ij}$ is the magnetic charge on site $i$. $\eta_i = 1\,(-1)$ if $i$ is an A (B) site.

The ground state manifold of $H_\mathrm{CSI}$ consists of states in which the magnetic charge vanishes everywhere: $Q_i=0,\,\forall i$.  Spinons are point defects carrying non-zero magnetic charge, which reside on diamond lattice sites. The excitation energy for a $Q=\pm1$ spinon is $J/2$.  As the energy only depends on magnetic charge, each degenerate manifold has a fixed number of spinons.

When quantum fluctuations set in, states belonging to the same degenerate manifold may tunnel into each other. As a result, the spinons can hop in the lattice via quantum spin flipping processes. Consider a $Q=1$ spinon at the site $i$. Of the four spins on the links emanating from $i$, three point toward $i$ and one points away from $i$. Flipping the spins that point toward $i$ moves the spinon to the neighbouring sites. However, flipping the spin that points away from $i$ creates a $Q=2$ spinon and a $Q=-1$ spinon, whereby the energy is increased by $J$ (Fig.\ref{pyro}b). When the energy scale of quantum fluctuations is small compared to $J$, one may neglect spinon creation processes. Thus, a $Q=1$ spinon must move against the direction of neighbouring spins. Likewise, a  $Q=-1$ spinon must move along the direction of the neighbouring spins~\cite{Chen1974,Jaubert2009,Jaubert2011,Petrova2015}.

We now write down the minimal model for spinon dynamics~\cite{Chen1974,Petrova2015}. We consider a lattice of $\mathcal{N}$ sites with periodic boundary conditions. Since we are interested in single spinon dynamics, we naturally wish to consider only one spinon.  However, a spin ice system with periodic boundary condition must be charge neutral. We therefore have to include a pair of $Q=\pm1$ spinons. The Hamiltonian in the 2-spinon Hilbert space is given by,
\begin{align}
H & = H_c+H_d  \nonumber\\
& = -\sum_{i\in A}\sum_{j\in N(i)} (c^{\dagger}_jc^{\phantom\dagger}_i S^{+}_{ij} + d^{\dagger}_jd^{\phantom\dagger}_i S^-_{ij}+h.c.).
\label{hamil}
\end{align}
Here, $H_c$ and $H_d$ are Hamiltonians describing the hopping of the $Q=1$ and $Q=-1$ spinon respectively, and we have scaled the spinon hopping amplitude to $1$. The first summation in the second line is over all A sublattice sites. $N(i)$ denotes the nearest neighbours of $i$. $c^\dagger_i (d^\dagger_i)$ creates a spinon at $i$, whereas $c_i(d_i)$ destroys a $Q=1(-1)$ spinon at $i$. Spinon operators belonging to different sites commute. We impose the hard-core condition that each site can be occupied by at most one spinon, $c^\dagger_ic^{\phantom\dagger}_i+d^\dagger_i d^{\phantom\dagger}_i\le 1$. $S^\pm_{ij}$ are spin raising and lowering operators.

Because the location of spinons and the spin configuration are not independent, we must impose the following magnetic charge condition as a constraint, $Q _i = c^\dagger_ic^{\phantom\dagger}_i - d^\dagger_id^{\phantom\dagger}_i$,
where $Q_i$ is the previously defined magnetic charge at site $i$. Note Eq.~\eqref{hamil} preserves this constraint. 

We now seek the approximate ground state of Eq.~\eqref{hamil} when $\mathcal{N}$ is large. To this end, we consider the column sum of Hamiltonian matrix elements, $R_\alpha  = \sum_{\beta} \langle \beta |H|\alpha\rangle$. $\alpha$ and $\beta$ label the basis states of the 2-spinon Hilbert space. $R_\alpha$ counts the number of flippable spins in $\alpha$. If the spinons are not in each other's proximity, there are 6 flippable spins. Thus, $R_\alpha = -6$. If the spinons occupy nearest-neighbour sites in $\alpha$, some of their hopping processes are blocked by the hard-core constraint, and hence $R_\alpha =  -4$ (Fig.~\ref{pyro}a). However, the fraction of such states in the Hilbert space is of order $1/\mathcal{N}$, which is negligible for large $\mathcal{N}$. Thus, the column sum of $H$ is a constant if we neglect rare spinon scattering events, which implies the following is an approximate eigenstate of Eq.~\eqref{hamil},
\begin{align}
|\Psi\rangle = \frac{1}{\sqrt{\Omega}}\sum_{\alpha}|\alpha\rangle.
\label{gs}
\end{align} 
$\Omega$ is the Hilbert space dimension. $H_c|\Psi\rangle \approx H_d|\Psi\rangle \approx -3|\Psi\rangle$. Provided that $H$ is ergodic, the Perron-Frobenius theorem implies Eq.~\eqref{gs} is the ground state~\cite{Barkema1998,Supp,Rokhsar1988}.

Having established the approximate ground state of Eq.~\eqref{hamil}, we move on to discuss the spinon dynamics. Ideally, one would like to study the spinon propagator. However, the usual single-particle propagator of a spinon is not gauge invariant, and hence its behaviour may depend on the gauge choice. Instead, we characterize the spinon dynamics through the time-dependent position correlation function. Without loss of generality, we consider the $Q=1$ spinon,
\begin{align}
C_{ij}(\tau) &\equiv \langle \mathcal{N} |e^{\tau H}c^\dagger_j c^{\phantom\dagger}_j e^{-\tau H} c^\dagger_i c^{\phantom\dagger}_i | \mathcal{N} \rangle/(1/\mathcal{N}) \nonumber\\
& \approx  \mathcal{N}e^{-3\tau} \langle \Psi | c^\dagger_j c^{\phantom\dagger}_j e^{-\tau H_c} c^\dagger_ic^{\phantom\dagger}_i | \Psi \rangle.
\label{pospos}
\end{align}
$C_{ij}$ measures the probability of finding the spinon on site $j$ at time $\tau$ given that it was found on site $i$ at time 0. $1/\mathcal{N}$ is the probability of observing the spinon on site $i$. $|\mathcal{N}\rangle$ is the exact ground state for finite $\mathcal{N}$. We have performed the Wick rotation, $it \to \tau$.  In the second line, we take the $\mathcal{N}\to \infty$ limit. In this limit, the two spinons are almost never in each other's proximity. Therefore, $[c^\dagger_ic^{\phantom\dagger}_i,H_d]\approx 0$ and $[H_c,H_d]\approx 0$.

\begin{figure}
\includegraphics[width=\columnwidth]{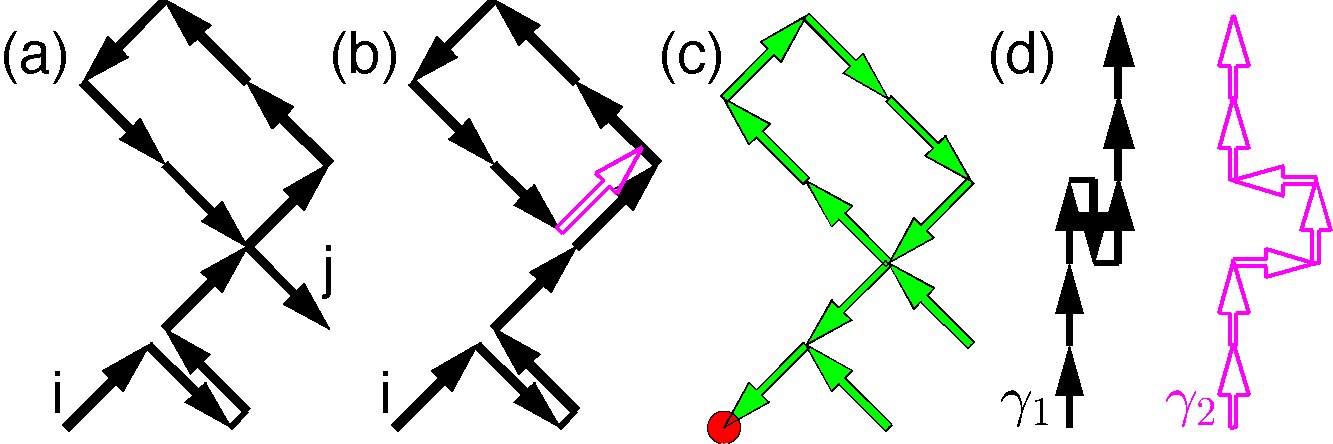}
\caption{(Colour online) (a) Projected view of a spinon walk $\gamma$. The rectangular loop represents a minimal-length loop in the diamond lattice. (b) A walk with zero weight. In the last step (magenta open arrow), the spinon traverses the link in the same direction as it did last time. (c) The trail of the walk in (a). Arrows indicate the spin orientation in the initial state. The filled red circle shows the initial position of the $Q=1$ spinon. (d) Two spinon walks $\gamma_1$ and $\gamma_2$ of the same length. $W_{\gamma_1}>W_{\gamma_2}$.}
\label{walk}
\end{figure}

We express Eq.~\eqref{pospos} as a path integral to make the relation between $C_{ij}(\tau)$ and the spinon walk explicit:
\begin{align}
C_{ij}(\tau) = e^{-3\tau}\sum_{\gamma}(\delta\tau)^{L}W_\gamma,
\label{pathint}
\end{align}
where the summation is over all $Q=1$ spinon walks $\gamma$ that starts from site $i$ at time $0$ and arrives at site $j$ at time $\tau$ (Fig.\ref{walk}a). At each step, the spinon either hops to neighbouring sites or stays at the same site within the time interval $\delta\tau$. $L$ is the total number of spinon hops. Note the $Q=-1$ spinon doesn't contribute as its position does not evolve under $H_c$.

The weighing factor $W_\gamma$ is defined as,
\begin{align}
W_\gamma &\equiv \mathcal{N}\langle \Psi| P_{j}S^\pm_{j,i_{L-1}}\cdots S^\pm_{i_2,i_1}S^\pm_{i_1,i} P_i|\Psi\rangle\nonumber \\
& =  \frac{\mathcal{N}}{\Omega}\sum_{\alpha,\beta}\langle \beta |P_{j}S^\pm_{j,i_{L-1}}\cdots S^\pm_{i_2,i_1}S^\pm_{i_1,i} P_{i}|\alpha\rangle.
\label{defweight}
\end{align}
$ii_1$, $i_1i_2\cdots i_{L-1}j$ stand for the time-ordered sequence of links traversed by the $Q=1$ spinon in the walk $\gamma$. $S^{\pm}_{ij}$ is either the raising or lowering operator, depending on the hopping direction (Eq.~\eqref{hamil}). $P_i$ projects onto basis states with the spinon located at site $i$. The second equality follows from Eq.~\eqref{gs}. $W_\gamma$ reflects the fact that the spinon hopping must be accompanied by successive spin flips. Had $W_\gamma$ been absent, Eq.~\eqref{pathint} would be identical to the path integral for a free particle.

We proceed to simplify Eq.~\eqref{defweight}. To this end, we consider a link $ab$, where $a\in A$ and $b\in B$. When the spinon hops from $a$ to $b$, the spin $S^z_{ab}$ flips from $-1/2$ to $1/2$. Next time the spinon traverses the link $ab$, it must hop from $b$ to $a$ so that $S^z_{ab}$ flips back to $-1/2$. Otherwise, the matrix element is zero. Thus, if the spinon is to pass a link multiple times, it must traverse the link in alternating directions. We denote the set of walks that obey this condition by $\Gamma$. $W_\gamma = 0$ if $\gamma\not\in\Gamma$ (Fig.~\ref{walk}b).

We then rearrange the product of spin operators, $S^\pm_{i_L,j_L}\cdots S^\pm_{i2,j2}S^\pm_{i1,j1} = \prod_{ab\in\gamma}(S^\pm_{ab}S^\pm_{ab}\cdots S^\pm_{ab})$, where the product is over the \emph{trail} of $\gamma$, which we define as the set of \emph{distinct} links traversed by the spinon. As $\gamma\in\Gamma$, $S^+_{ab}$ and $S^-_{ab}$ appear alternatively. Thus,
\begin{align}
\sum_{\sigma'_{ab}}\langle \sigma'_{ab} | \cdots S^\pm_{ab}S^\mp_{ab}S^\pm_{ab} |\sigma_{ab}\rangle = \delta_{\sigma_{ab},\mp1/2}.
\label{weightprep1}
\end{align}
Here $\sigma_{ab}$ is the eigenstate of $S^z_{ab}$. The upper case corresponds to $S^{+}_{ab}$ appearing at the rightmost position in the string. In this case, the $Q=1$ spinon hops from $a$ to $b$ when it traverses the link $ab$ for the first time. The spin $S^z_{ab}$ must point from $b$ to $a$ in the initial state, which is reflected in the Kronecker delta $\delta_{\sigma_{ab},-1/2}$. Similarly, the lower case corresponds to the spinon hopping from $b$ to $a$ upon the first traversal, and the Kronecker delta $\delta_{\sigma_{ab},1/2}$ ensures the spin $S^z_{ab}$ points from $a$ to $b$. 

Combining Eqs.~\eqref{defweight} and \eqref{weightprep1}, we find
\begin{align}
W_\gamma=
\frac{\sum_{\alpha}P_{i}(\alpha)\prod_{ab\in\gamma}\delta_{\sigma_{ab},\pm1/2}}{\sum_{\alpha}P_i(\alpha)}\quad (\gamma\in\Gamma).
\label{weight}
\end{align}
$P_i(\alpha)$ is the diagonal matrix element of $P_i$. $P_i(\alpha)=1$ if the $Q=1$ spinon is at site $i$ in the initial state $\alpha$, and $P_i(\alpha)=0$ otherwise. The Kronecker deltas select the initial states in which $\sigma_{ab}$, the spins residing on the trail of $\gamma$, point against the direction of the first traversal (Fig.~\ref{walk}c). We have replaced $\Omega/\mathcal{N}$ by $\sum_{\alpha}P_i(\alpha)$, which follows from the lattice translation symmetry.

Eqs.~\eqref{pathint} and \eqref{weight} show that the spinon acquires a memory after integrating out background spins. In particular, $W_\gamma$ is entropic in nature. The numerator in Eq.~\eqref{weight} counts the number of initial states that are compatible with the walk $\gamma$, whereas the denominator counts the total number of initial states. The entropic nature of $W_\gamma$ implies that the spinon prefers to retrace the links it traversed before. To see this, we compare two walks $\gamma_1$ and $\gamma_2$ of the same length (Fig.~\ref{walk}d). The spinon retraces a link twice in $\gamma_1$, and hence the trail of $\gamma_1$ has fewer links than $\gamma_2$. Since fewer spins have their orientation fixed in $\gamma_1$, more initial states are compatible with $\gamma_1$ than $\gamma_2$. Thus, $\gamma_1$ has larger contribution to Eq.~\eqref{pathint}.

\begin{figure}
\includegraphics[width=\columnwidth]{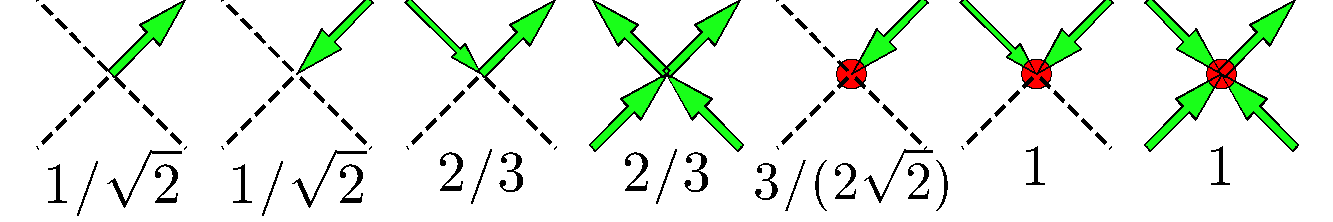}
\caption{(Colour online) Vertex configurations and their weights in Pauling approximation. The $Q=1$ spinon is shown as a filled red circle. Links that are not traversed by the spinon are shown as dashed lines. Permutations of a vertex configuration have the same weight.}
\label{pauling}
\end{figure}

In lieu of exact formulae, we estimate $W_\gamma$ by generalizing the Pauling approximation (PA)~\cite{Pauling1935}. The PA treats the spin correlation within a vertex, i.e.~the group of four spins surrounding a lattice site, exactly and neglects long-range spin correlations. We find~\cite{Supp},
\begin{align}
W_\gamma \approx  \prod_{a}w_a \quad (\gamma\in\Gamma) . \label{paul_aprx}
\end{align}
Here, $w_a$ is the contribution from the vertex at site $a$. The values of $w_a$ are tabulated in Fig.~\ref{pauling}. The product is over all sites $a$ in the trail. To assess the accuracy of PA, we perform an exact enumeration of Eq.~\eqref{weight} in a 40-spin cluster. For trails of up to 7 steps, the values of $W_\gamma$ from the two approaches differ by less than $1\%$~\cite{Supp}. We are thus confident that the PA should be reliable.

\begin{figure}
\includegraphics[width=\columnwidth]{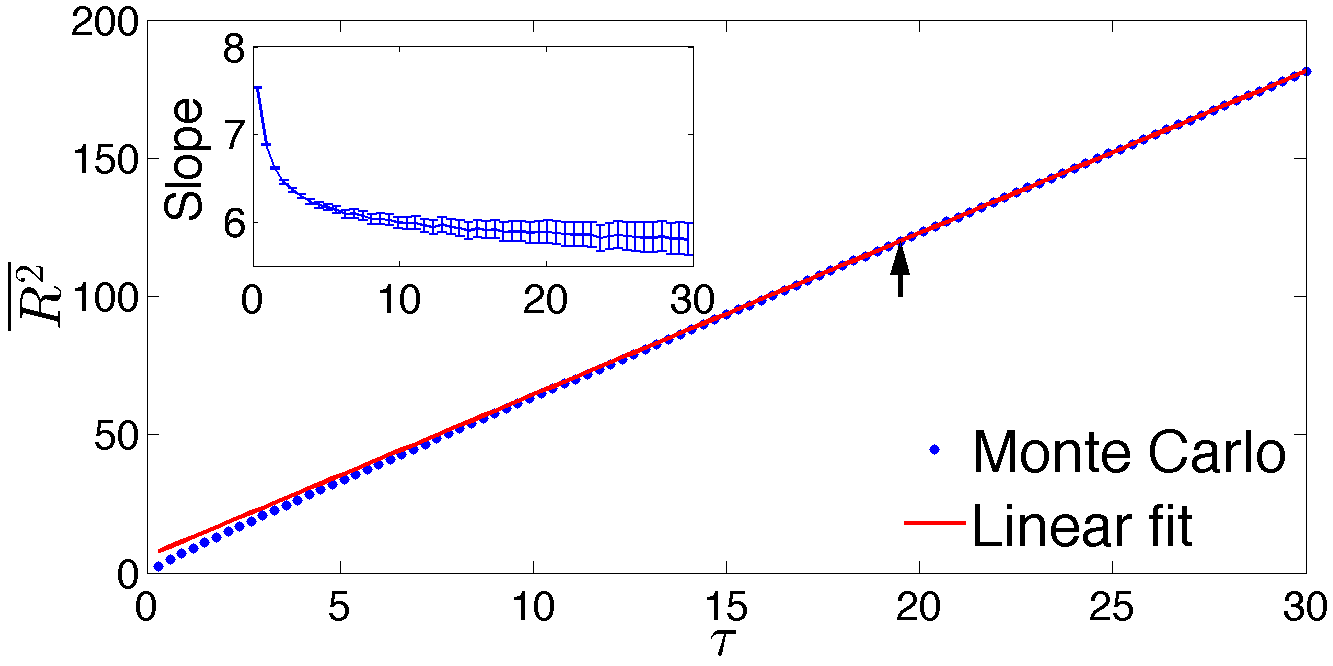}
\caption{(Colour online) $\overline{R^2}$ as a function of imaginary time $\tau$. Error bars are smaller than the symbols. Arrow indicates the starting point of linear fit. Inset shows the slope of $\overline{R^2}(\tau)$.}
\label{r2}
\end{figure}

Equipped with Eq.~\eqref{paul_aprx}, we simulate the path integral Eq.~\eqref{pathint} via a continuous time Monte Carlo procedure \cite{Henley2004, Svljuaasen2005, Supp}. We first study the mean squared displacement $\overline{R^2}(\tau) \equiv \sum_{j}|\mathbf{R}_{ij}|^2C_{ij}(\tau)$ (Fig.~\ref{r2}). $\mathbf{R}_{ij}$ is the vector pointing from $i$ to $j$. The data show the slope of $\overline{R^2}$ monotonically decreases, reflecting the retracing nature of the spinon walk. For large $\tau$, $\overline{R^2}$ approaches linear behaviour. Recall that the Schr\"{o}dinger equation becomes a diffusion equation after the Wick rotation. In particular, $\overline{R^2} = 3\tau/m$ for a free particle with mass $m$ moving in three dimensions. Thus, the data indicate that the spinon propagates as a free massive particle at low energy. To estimate the spinon effective mass $m_\mathrm{eff}$, we fit $\overline{R^2}$ to a linear function and find $m_\mathrm{eff}\approx 0.51$,  which is slightly larger than twice that of the tight-binding effective mass $m_\mathrm{tb} = 0.25$.

\begin{figure}
\includegraphics[width=\columnwidth]{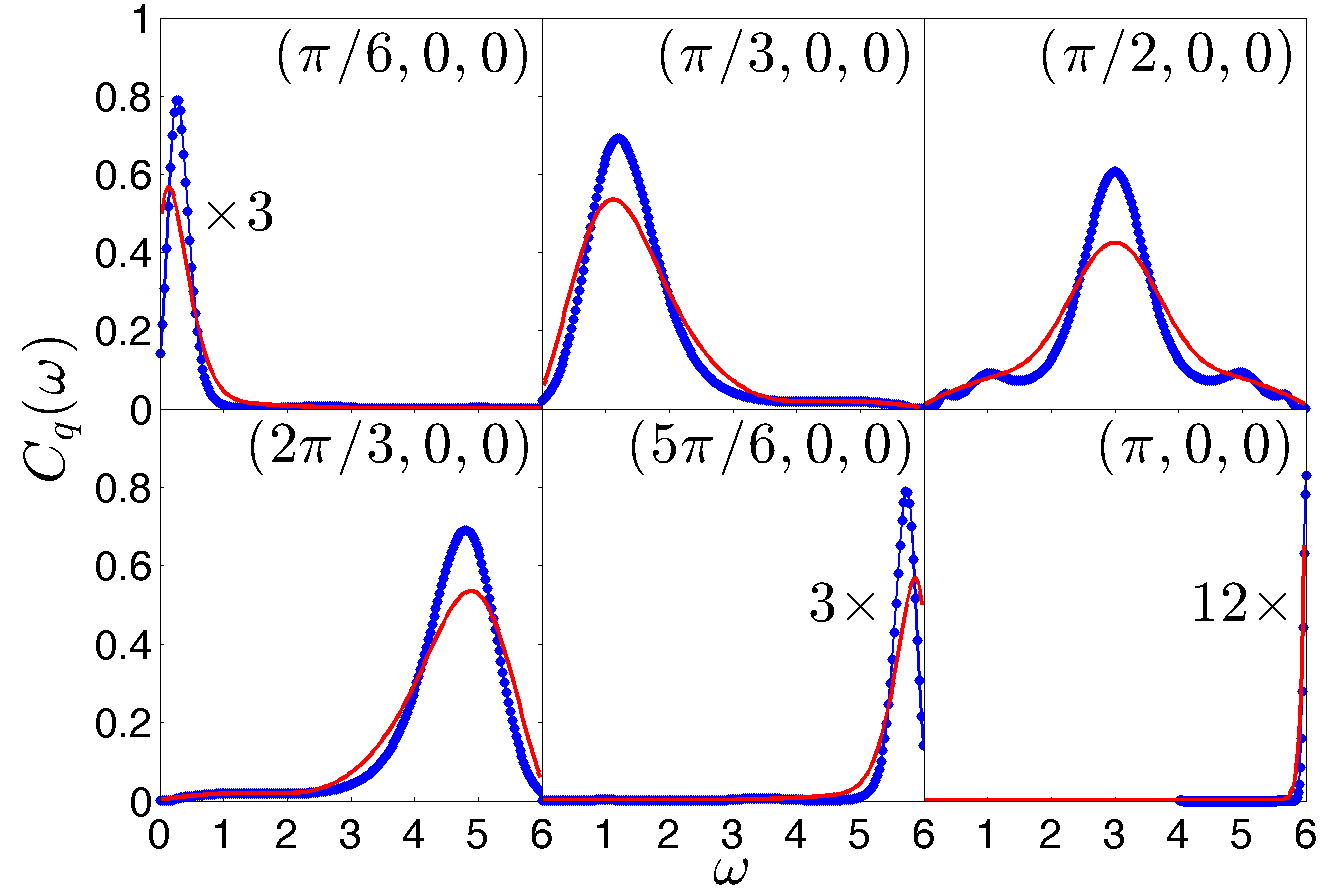}
\caption{(Colour online) $C_\mathbf{q}(\omega)$ along the high symmetry direction $\mathbf{q}\parallel (1,0,0)$. The spectra are constructed from maximum entropy analytic continuation (blue dots) and moment expansion (red line). Note the data for some $\mathbf{q}$ are rescaled.}
\label{spectrum}
\end{figure}

To corroborate the above picture, we study $C_\mathbf{q}(\omega)$, the Fourier transform of $C_{ij}(t)$ in space and real time $t$. The spinon band bottom and top are located at the energy $-3$ and 3 respectively. Thus, $C_\mathbf{q}(\omega)$ is non-zero only in the frequency window $\omega \in [0,6]$. Fig.~\ref{spectrum} presents $C_\mathbf{q}(\omega)$ along the high symmetry direction $(q,0,0)$. $C_\mathbf{q}(\omega)$ is constructed from the imaginary time data $C_\mathbf{q}(\tau)$ through the maximum entropy analytic continuation~\cite{Jarrell1996,Mead1984,Supp}. The spectra at $q<\pi/2$ is obtained from $\pi-q>\pi/2$ by using the sublattice symmetry $C_\mathbf{q}(\omega)=C_{\mathbf{Q}-\mathbf{q}}(6-\omega)$, where $\mathbf{Q}=(\pi,0,0)$. We also use the first 16 power moments to construct the spectrum~\cite{Weisse2006,Supp}.

Our model conserves the spinon number, which implies $C_{\mathbf{q}=0}(\omega) = \delta(\omega)$ (not shown). We thus expect $C_{\mathbf{Q}}(\omega)=\delta(\omega-6)$ from the sublattice symmetry. The spectrum at $q = \pi$ indeed is a sharp peak at $\omega = 6$. For $q$ near $0$ or $\pi$, the peak remains sharp. This can be compared to a free particle in the continuum, whose $C_{\mathbf{q}}(\omega)$ is a delta function located at the particle kinetic energy. Our data suggest that the spinon behaves as a quasiparticle near the band bottom and top. The peak broadens as $q$ approaches $q=\pi/2$, at which point we observe a broad main peak at $\omega=3$ and two satellite peaks, indicating the spinon propagation is much less coherent at the band centre.

The emergence of near-free particle behaviour from the intricate interplay between the spinon and the spin background comes as a surprise, which we heuristically understand as a consequence of spinon retracing. The spinon strongly interacts with itself when it travels in a loop and revisits a link. The interaction comes as a hard constraint on the direction of traversal (Fig.\ref{walk}b). However, the spinon prefers retracing its previous steps, which reduces the probability of the revisiting events. Such interaction thus becomes ineffective. Furthermore, since retracing makes it harder to propagate through the lattice, the spinon acquires a larger effective mass than its tight-binding counterpart.

Our study of the minimal model Eq.~\eqref{hamil} provides a starting point for understanding the dynamics of QSI. The model is applicable when the spinon interaction can be neglected and the quantum fluctuation energy scale $h$ is small compared to spin ice exchange energy $J$. Beyond the minimal model, many interesting questions remain open. In some QSI materials, the spin dipolar interaction may be important. It will induce a Coulomb force between the spinons, which gives the possibility of spinon bound states~\cite{Castelnovo2008,Petrova2015}. Furthermore, more careful modelling is required when $h$ and $J$ are comparable,  such as in the material Yb$_2$Ti$_2$O$_7$~\cite{Ross2011}. Finally, our study focuses on single-spinon dynamics. A theory of many-spinon dynamics would be needed to fully address experiments probing QSI dynamics at hydrodynamic time scales~\cite{Pan2016,Tokiwa2016}. 

\begin{acknowledgments}

We are grateful to John Chalker, Michel Gingras, Stefanos Kourtis, and Jeff Rau for discussions, and to Claudio Castelnovo and Oleg Tchernyshyov for critical reading of our manuscript. This research was supported by NSERC (R.G.M.), the Canada Research Chair program (R.G.M.), and the John Templeton Foundation (R.G.M. and J.C.). Research at Perimeter Institute is supported by the Government of Canada through Industry Canada and by the Province of Ontario through the Ministry of Research and Innovation.

\end{acknowledgments}

\bibliographystyle{apsrev4-1}
\bibliography{spinon}

\onecolumngrid
\appendix

\section{Ergodicity of the spinon Hamiltonian}

We argue that the Hamiltonian
\begin{align}
H  =  -\sum_{i\in A}\sum_{j\in N(i)} (c^{\dagger}_jc_i S^{+}_{ij} + d^{\dagger}_jd_i S^{-}_{ij}+h.c.).
\label{supp_hamil}
\end{align}
is ergodic in the 2-spinon Hilbert space. In other words, one can reach any basis state $|\beta\rangle$ by repeatedly acting the Hamiltonian $H$ on any other basis state $|\alpha\rangle$. We take the lattice to be finite and impose periodic boundary conditions.

To this end, we consider the transition graph from $|\alpha\rangle$ to $|\beta\rangle$~\cite{Barkema1998}. The transition graph is constructed as follows: one compares the state $\beta$ and $\alpha$. If a spin $S^z_{ij}$ is flipped in $\beta$ with respect to state $\alpha$, we colour the link $ij$. The transition graph $G$ is made of the coloured links.

In the absence of spinons, the transition graphs $G$ consists of loops. In this case, the zero magnetic charge condition $Q_i=0$ is enforced on every vertex. To preserve the $Q_i=0$ condition, spins belonging to the vertex $i$  must be flipped in pairs. Thus, the degree of each site in the transition graph $G$ is an even number, which implies that $G$ is made of loops. Furthermore, the spins in the same loop point along the same direction in state $\alpha$.

If the spinons are present, the transition graph $G$ consists of both loops and open strings. The end points of the open strings coincide with the location of spinons in states $\alpha$ and $\beta$. The magnetic charge $Q_i=\pm1$ on lattice sites occupied by spinons. Thus, in the transition graph $G$, the degree of spinon sites is an odd number, which implies these sites are end points of open strings. The spins belonging to the same open string all point along the same direction in state $\alpha$.

In particular, when there are two spinons in the system, four sites of the transition graph are occupied by the spinons. We label the spinon sites as $x_\alpha$, $x_\beta$, $y_\alpha$, and $y_\beta$. $x_\alpha$ and $y_\alpha$ are the location of $Q=1$ and $Q=-1$ spinons in state $\alpha$, respectively. By the same token, $x_\beta$ and $y_\beta$ are the location of $Q=1$ and $Q=-1$ spinons in state $\beta$, respectively. The topology of open strings thus fall into two kinds. In the first kind (Fig.\ref{supp_openstring}a), $x_\alpha$ and $x_\beta$ are connected by an open string, and so are $y_\alpha$ and $y_\beta$. In the second kind (Fig.\ref{supp_openstring}d), $x_\alpha$ and $y_\alpha$ are connected, and $x_\beta$ and $y_\beta$ are connected. Note that it is impossible to connect $x_\alpha$ to $y_\beta$, i.e. to connect a $Q=1$ spinon to a $Q=-1$ spinon, as the magnetic charge of a spinon doesn't change.

The Hamiltonian Eq.~\eqref{supp_hamil} describes the hopping of spinons. Demonstrating that Eq.~\eqref{supp_hamil} is ergodic amounts to showing that, for any transition graph $G$, one can account for the corresponding spin flips by spinon moves. It is sufficient to consider three elementary transition graphs: (a) $G$ is a loop; (b) $G$ consists of two open strings of the first kind; (c) $G$ consists of two open strings of the second kind.  All other transitions graphs are composition of these three elementary cases.

\begin{figure}
\includegraphics[width=\textwidth]{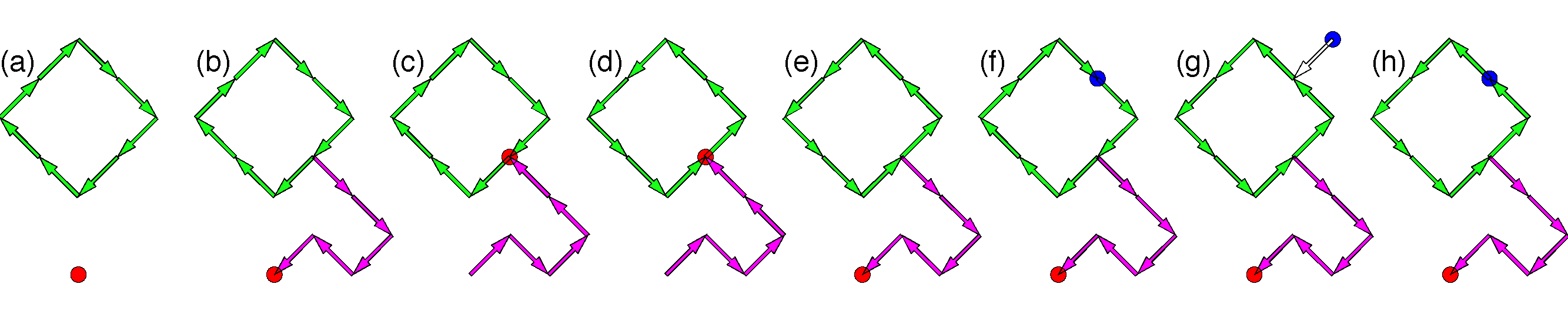}
\caption{(a) Projected view of a loop in the transition graph from the state $\alpha$ to $\beta$. Arrows show the orientation of the spins in the state $\alpha$. Red circle denotes a $Q=1$ spinon. (b) The path $\gamma$ (magenta arrows) connects the  spinon to a site belonging to the loop. (c) Transporting the spinon along the path $\gamma$ flips all the spins along the path. (d) Moving the spinon around the loop flips all the spins in the loop. (e) Transporting the spinon back to its original location along $\gamma$ restores the spin orientation in $\gamma$. (f) A $Q=-1$ spinon (blue circle) resides on the loop. (g) After the $Q=-1$ spinon being moved away from the loop, spins in the loop are flipped by following (b)(c)(d)(e). (h) The $Q=-1$ spinon is brought back to its original location.}
\label{supp_loop}
\end{figure}

To proceed, we need the following lemma: In a sufficiently large diamond lattice with periodic boundary conditions, the spinons can reach any lattice site $i$ starting from any state $\alpha$ by successive spin flips. We provisionally accept the lemma without proof but provide arguments in its favour at the end of this section.

We first consider elementary case (a). In this case, we may reach state $\beta$ from $\alpha$ by flipping all spins in the loop (Fig.\ref{supp_loop}a). We can do this in three steps. First, we move a $Q=1$ spinon from $x_\alpha$ to $i$, a site belonging to the loop, along a path $\gamma$ (Fig.\ref{supp_loop}b\&{}c). This is possible thanks to the lemma. In particular, the spins on $\gamma$ all point from $i$ to $x_\alpha$ in state $\alpha$. After transporting the spinon along $\gamma$, the spins now all point from $i$ to $x_\alpha$. Next, the spinon is moved around the loop and thereby flip all the spins in the loop (Fig.\ref{supp_loop}d). Last, the spinon is transported back from $i$ to $x_\alpha$ along the path $\gamma$ (Fig.\ref{supp_loop}e). As a result, the states of spins on $\gamma$ are restored.

We need to modify the above operations slightly if the $Q=-1$ spinon resides on the loop (Fig.\ref{supp_loop}f). In this case, we need first move the $Q=-1$ spinon one step away from the loop (Fig.\ref{supp_loop}g), move the $Q=1$ spinon around the loop, and then move the $Q=-1$ spinon back to its original location (Fig.\ref{supp_loop}h).

\begin{figure}
\includegraphics[width=\textwidth]{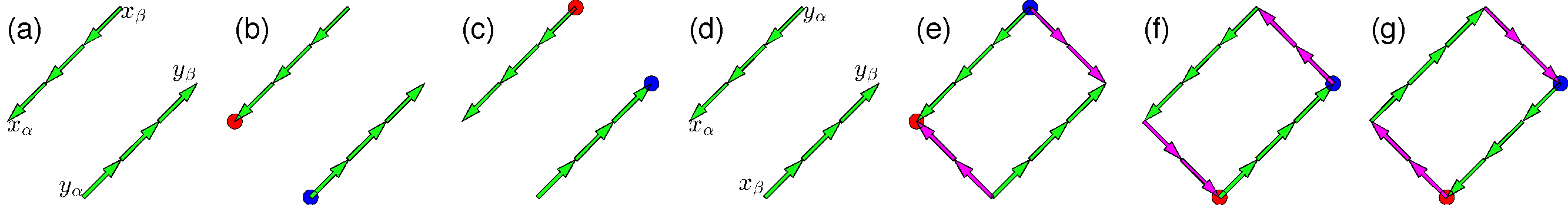}
\caption{(a) Projected view of the open strings of the first kind in transition graph from the state $\alpha$ to $\beta$. Arrows show the orientation of spins in the state $\alpha$. $x_\alpha(x_\beta)$ and $y_\alpha(y_\beta)$ are the locations of $Q=1$ and $-1$ spinons in the state $\alpha (\beta)$. (b) Spin orientations in the state $\alpha$. Red and blue circles show the location of $Q=1$ and $-1$ spinons respectively. (c) Transporting the spinons along the open strings flip all spins in the transition graph. (d) Projected view of the open strings of the second kind in transition graph from the state $\alpha$ to $\beta$ . (e) The string $\gamma_3$ (magenta arrows) connects $x_\alpha$ to $x_\beta$,  and the string $\gamma_4$ (magenta arrows) connects $y_\alpha$ to $y_\beta$. (f) Transporting spinons along $\gamma_3$ and $\gamma_4$ flips all spins on $\gamma_3$ and $\gamma_4$. $\gamma_1$,$\gamma_2$,$\gamma_3$, and $\gamma_4$ form a loop. Spins in this loop all point in the same direction. (g) Following the operations in Fig.\ref{supp_loop}, one can flip all the spins in the loop. The spins on $\gamma_3$ and $\gamma_4$ are thereby restored to their original orientation.}
\label{supp_openstring}
\end{figure}

We then consider the elementary case (b) (Fig.\ref{supp_openstring}a). In this case, one open string $\gamma_1$ connects the point $x_\alpha$ and $x_\beta$, and the other open string $\gamma_2$ connects $y_\alpha$ and $y_\beta$. We can reach $\beta$ from $\alpha$ by first moving the $Q=1$ spinon from $x_\alpha$ to $x_\beta$ along $\gamma_1$ and then moving $Q=-1$ spinon from $y_\alpha$ to $y_\beta$ along $\gamma_2$ (Fig.\ref{supp_openstring}b\&{}c).

Finally, we consider the elementary case (c) (Fig.\ref{supp_openstring}d). In this case, the one open string $\gamma_1$ connects the point $x_\alpha$ and $y_\alpha$, and the other open string $\gamma_2$ connects $x_\beta$ and $y_\beta$. In the state $\alpha$, all the spins on $\gamma_1$ point from $y_\alpha$ to $x_\alpha$, and all the spins on $\gamma_2$ point from $x_\beta$ to $y_\beta$. We can reach $\beta$ from $\alpha$ in two steps. In the first step, we transport the $Q=1$ spinon from $x_\alpha$ to $x_\beta$ along some path $\gamma_3$ and the $Q=-1$ spinon from $y_\alpha$ to $y_\beta$ along $\gamma_4$ (Fig.\ref{supp_openstring}e\&{}f). This operation is possible thanks to the lemma. As a result, all the spins on $\gamma_3$ now point from $x_\alpha$ to $x_\beta$, and all the spins on $\gamma_4$ now point from $y_\beta$ to $y_\alpha$. In the next step, we observe that $\gamma_{1}$, $\gamma_2$, $\gamma_3$, and $\gamma_4$ form a loop. Thus, we move the $Q=1$ spinon around this loop following the same procedure in the elementary case (a) (Fig.\ref{supp_openstring}(g)). As a result, the spins on $\gamma_1$ and $\gamma_2$ are flipped, whereas spins on $\gamma_3$ and $\gamma_4$ are restored to their original state in $\alpha$.

To sum up, we have shown that the Hamiltonian Eq.\ref{supp_hamil} is ergodic given the lemma. We now argue that the lemma is indeed true. Without loss of generality, we consider the $Q=1$ spinon. It is sufficient to show that, in any basis $\alpha$, there exists a string of spins that connect $x_\alpha$, the location of the $Q=1$ spinon, and an arbitrary lattice point $i$ and that all the spins on the string point toward the spinon.

\begin{figure}
\includegraphics[width = 0.4\textwidth]{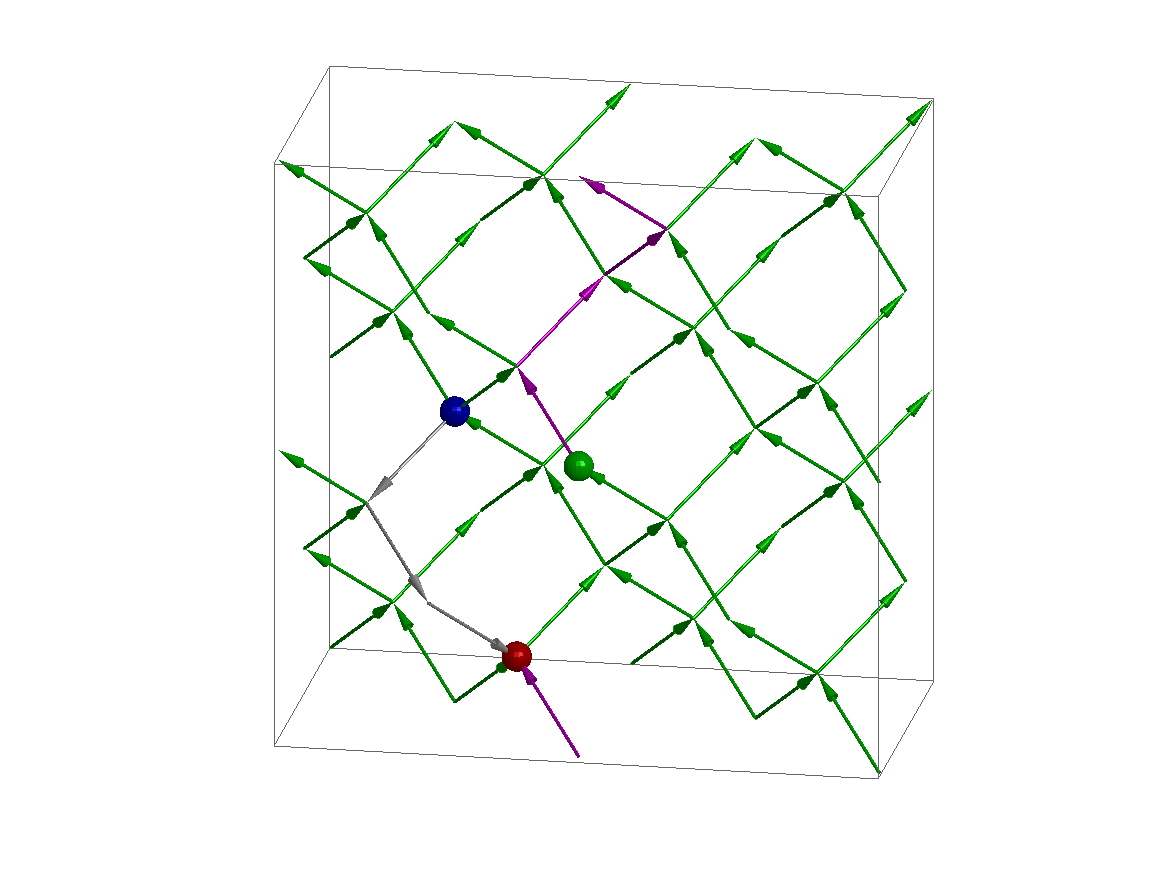}
\caption{A maximally polarized state. All spins (green arrows) point upward except for a string of downward spins (grey arrows). Periodic boundary condition is imposed between on top and the bottom plane. Red and blue balls show the location of $Q=1$ and $-1$ spinons respectively. Green ball shows an arbitrary site $i$ in the diamond lattice. A path (magenta arrows) connects the $Q=1$ spinon to the site $i$. All spins on the path points toward the spinon.}
\label{supp_path}
\end{figure}

Let us first consider a maximally polarized state $\alpha_0$ (Fig.\ref{supp_path}). In $\alpha_0$, all spins point upward except for a string of downward spins connecting the $Q=1$ spinon and the $Q=-1$ spinon. It is easy to see that there exists a string of spins connecting the $Q=1$ spinon and an arbitrary lattice point $i$. The spins on the string all point from $i$ to $x_{\alpha_0}$. We denote this string by $\gamma$. Note that the periodic boundary condition is crucial. Otherwise, the spinon can only move downward and the proposition is false.

We now consider a generic state $\alpha$ in which the spinons are located at the same sites as in $\alpha_0$. The transition graph from $\alpha_0$ to $\alpha$ consists of only loops as the spinons don't move. In other words, one can reach the state $\alpha$ from the state $\alpha_0$ by flipping spins around loops. Such loop flipping operations may deform the said string $\gamma$, but they will never break it. Thus, we see that the said string $\gamma$ also exists in the state $\alpha$.

We have also performed numerical checks on the ergodicity of Hamiltonian Eq.~\eqref{supp_hamil} on a two-dimensional square lattice. Numerical results show that $H$ is ergodic on $2\times3$ and $3\times3$ lattices.

\section{Pauling approximation}

The weighting factor $W_\gamma$ is given by,
\begin{align}
W_\gamma=
\frac{\sum_{\alpha}P_{i}(\alpha)\prod_{\langle ab\rangle}\delta_{\sigma_{ab},\pm1/2}}{\sum_{\alpha}P_i(\alpha)}\quad (\gamma\in\Gamma).
\label{supp_weight}
\end{align}
We estimate Eq.~\eqref{supp_weight} by generalizing the Pauling approximation. 

In the Pauling approximation, the correlation within a vertex, i.e. the four spins around a site, is treated exactly, but long-range spin correlations are essentially neglected. We first consider the denominator of the Eq.~\eqref{supp_weight}:
\begin{align}
\sum_{\alpha}P_{i}(\alpha)  = \sum_{\sigma} \prod_a V_a (\sigma_{a1},\sigma_{a2},\sigma_{a3},\sigma_{a4}) = 2^{2\mathcal{N}}\langle \prod_a V_a (\sigma_{a1},\sigma_{a2},\sigma_{a3},\sigma_{a4}) \rangle.
\end{align}
Here $\sigma_{a1},\sigma_{a2},\sigma_{a3},\sigma_{a4}$ are the four spins surrounding the site $a$, or the vertex at $a$. $V_a$ is the interaction between the four spins, which enforces the magnetic charge conditions. Specifically,
\begin{align}
V_a(\sigma_{a1},\sigma_{a2},\sigma_{a3},\sigma_{a4}) = \left\{\begin{array}{cc}
1 & \textrm{if }\sigma_{a1,a2,a3,a4}\textrm{ obey the magnetic charge condition on }a\\
0 & \textrm{otherwise}
\end{array}\right. .
\end{align}
Recall the spinon resides on site $i$. On this site, the condition $Q_a = 1$ is enforced. When $a\neq i$, the condition $Q_a=0$ is enforced. $\sum_\sigma$ is the summation over all spin states. $\langle\cdots\rangle$ stands for the average in the ensemble where each spin is independent and takes value $\pm1/2$ with equal probability. $2\mathcal{N}$ is the total number of spins. In the Pauling approximation, the vertices are treated as independent,
\begin{align}
\sum_{\sigma}P_{i}(\sigma) \approx 2^{2\mathcal{N}}\prod_a \langle V_a\rangle.
\end{align}

\begin{figure}
\includegraphics[width = 0.4\textwidth]{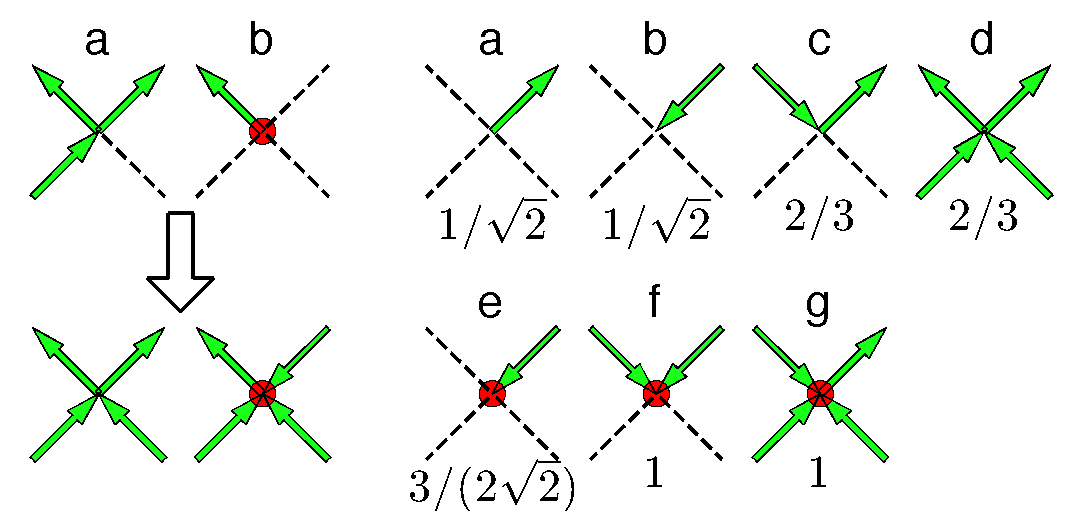}
\caption{Left: the spins on links that are not traversed by spinon are free to take two possible orientations in general (top row). However, thanks to the magnetic charge condition, the orientation of spins may be inferred (bottom row). Dashed lines denote links not traversed by the spinon. The filled red circle denotes a $Q=1$ spinon. Right: weights assigned to vertex configurations. Configurations with zero weights are not shown.}
\label{supp_pauling}
\end{figure}

By the same token, we can estimate the denominator of Eq.~\eqref{supp_weight}. Similar to the denominator, the numerator also counts the number of spin states that obey the magnetic charge conditions. Different from the denominator, the orientation of spins on trail, $\sigma_{ab}$, is fixed whereas the other spins are free to take two possible orientations $\pm1/2$. However, thanks to the magnetic charge conditions, the orientation of some spins, even though they do not belong to the trail, can be inferred. As an example, we consider the vertex configuration (a) shown in the left panel of Fig.\ref{supp_pauling}. There is no spinon present at this vertex, and therefore two spins much point toward the vertex and the other two spins must point away. Here the orientation of three spins are already known, which implies that the spin on the untraversed link (dashed line) must point toward the vertex. In this case, we extend the trail by including this new spin. Similar reasoning applies to another configuration (b) shown in the left panel of Fig.\ref{supp_pauling}.

After extending the trail, we can now estimate the numerator,
\begin{align}
&\sum_{\sigma}P_{i}(\sigma)\prod_{\langle ab\rangle}\delta_{\sigma_{ab},\pm1/2} = \langle \prod_a V_a (\sigma_{a1},\sigma_{a2},\sigma_{a3},\sigma_{a4}) \rangle'  \approx 2^{2\mathcal{N}-l}\prod_a \langle V_a \rangle'.
\end{align}
$\langle\cdots\rangle'$ stands for the average in the ensemble where $\sigma_{ab}$, the spins on trail of $\gamma$, take fixed value while the rest are independent and free. $l$ is the number of spins on trail. $2\mathcal{N}-l$ is the number of free spins. In the last step, we again use the Pauling approximation.

Combining the above results, the weighting factor $W(\gamma)$ takes the following form,
\begin{align}
W_\gamma &\approx 2^{-l}\prod_{a\in\gamma}\frac{\langle V_a\rangle'}{\langle V_a \rangle} = \prod_{a\in\gamma}2^{-l_a/2}\frac{\langle V_a\rangle'}{\langle V_a \rangle} \equiv \prod_{a\in\gamma}w_a.\quad (\gamma\in\Gamma)
\end{align}
$a\in\gamma$ stands for the sites visited by the spinon. The first equality follows from the fact that $\langle V_a\rangle=\langle V_a\rangle'$ if the vertex $a$ is not visited by the spinon.  $l_a$ is the number of traversed links in vertex $a$. In the second equality, we have used the fact that $\sum_a l_a  = 2l$. $w_a$ is the weight associated with the site $a$. $w_a$ only depends on the configuration of the vertex at $a$. Its values are tabulated in the right panel of Fig.\ref{supp_pauling}. Note that two configurations that are related by permuting the four spins of the vertex have the same weight.

To see how the results in the right panel of Fig.\ref{supp_pauling} are derived, we take the vertex configuration $(c)$ as an example. The weights for other vertex configurations can be derived along similar lines. The condition $Q=0$ is imposed on this vertex. On the one hand, if we do not specify the spin orientation at all, the vertex has 4 free spins, or $2^4$ states, 6 of which obey the $Q=0$ condition. Thus, $\langle V_a\rangle=3/8$. On the other hand, in the configuration $(c)$, the orientation of two spins are fixed, and the other two are free. The vertex can take 4 states, 2 of which obey the constraint. Thus, $\langle V_a\rangle' = 1/2$. Combining the two results, we find $w_a=2^{-l_a/2}\langle V_a\rangle'/\langle V_a\rangle=2/3$.

\begin{figure}
\includegraphics[width = 0.4\textwidth]{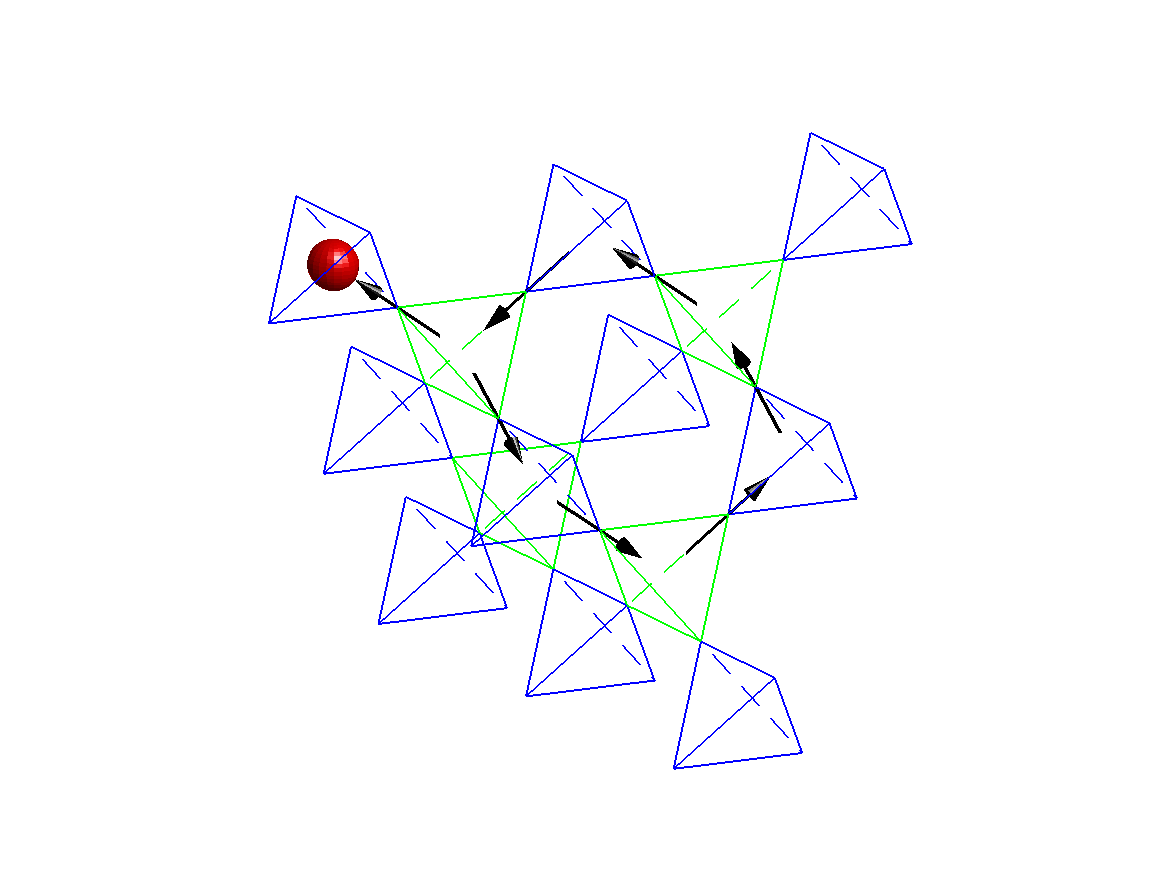}
\caption{The 40-spin cluster used for enumeration. The solid red ball denotes a $Q=1$ spinon. A trail of 7 spins (black arrows) is also shown.}
\label{supp_cluster40}
\end{figure}

To assess the accuracy of the Pauling approximation, we have also performed an exact enumeration of Eq.\ref{supp_weight} on a cluster of 40 spins with open boundary condition (Fig.\ref{supp_cluster40}). We take several trails with various length and shape. We define the relative error $\epsilon\equiv |W^\mathrm{enum}_\gamma-W^\mathrm{Pauling}_\gamma|/W^\mathrm{enum}_\gamma$. We find that, for trails of up to 7 steps, the values of $W_\gamma$ from the two different approaches differ by less than $1\%$. For details see Table~\ref{Pap}. Therefore, we are confident that the Pauling approximation should be reliable.

\begin{table}
\centering
\begin{tabular}{cccc}
\hline
\hline
trail length & $W^\mathrm{enum}_\gamma$ & $W^\mathrm{Pauling}_\gamma$ & $\epsilon$ \\
\hline
1 & 0.75 & 0.75 & 0\\
2 & 0.5 & 0.5 & 0\\
3 & 0.3333 & 0.3333 & 0\\
4a & 0.2230 & 0.2222 & $3.623\times 10^{-3}$ \\
4b & 0.2218 & 0.2222 & $1.821\times 10^{-3}$ \\
5 & 0.1479 & 0.1481 & $1.821\times 10^{-3}$ \\
6a & 0.09939 & 0.09876 & $6.323\times 10^{-3}$\\
6b & 0.09943 & 0.09876 & $6.672\times 10^{-3}$\\
6c & 0.09943 & 0.09876 & $6.672\times 10^{-3}$\\
7 & 0.06545 & 0.06584 & $5.944\times 10^{-3}$\\
\hline
\hline
\end{tabular}
\caption{Comparison between the exact enumeration and the Pauling approximation for different trails. $W^\mathrm{enum}_\gamma$ stands for the result from the enumeration whereas $W^\mathrm{Pauling}_\gamma$ stands for the result from Pauling estimate. The relative error $\epsilon \equiv |W^\mathrm{enum}_\gamma-W^\mathrm{Pauling}_\gamma|/W^\mathrm{enum}_\gamma$.}\label{Pap}
\end{table}

\section{Continuous time Monte Carlo}\label{supp_ctmc}

The position correlation function $C_{ij}(\tau)$ is written as,
\begin{align}
C_{ij}(\tau) \equiv e^{-3\tau}\mathcal{N}\langle\Phi|c^\dagger_j c_j e^{-\tau H_c}c^\dagger_i c_i|\Phi\rangle = e^{-3\tau}\frac{\langle\Phi|c^\dagger_j c_j e^{-\tau H_c}c^\dagger_i c_i|\Phi\rangle}{\langle\Phi|c^\dagger_i c_i|\Phi\rangle} = \frac{\langle\Phi|c^\dagger_j c_j e^{-\tau H_c}c^\dagger_i c_i|\Phi\rangle}{\langle\Phi| e^{-\tau H_c}c^\dagger_ic_i|\Phi\rangle}.
\label{supp_cdef}
\end{align}
The second equality follows from the fact that $\langle\Phi|c^\dagger_ic_i|\Phi\rangle = 1/\mathcal{N}$. The third equality follows from $H_c|\Phi\rangle = -3|\Phi\rangle$.

Following the derivation in the paper, we rewrite $C_{ij}(\tau)$ as a path integral,
\begin{align}
C_{ij}(\tau) = \frac{\sum_{\gamma:i\to j}W_\gamma(d\tau)^{L_\gamma}}{\sum_{\gamma:i}W_\gamma(d\tau)^{L_\gamma}}.
\label{supp_pathint}
\end{align}
In the numerator, the summation is over all walks $\gamma$ starting from site $i$ at time 0 and arriving at site $j$ at time $\tau$. The time lapse of each step is $\delta\tau$. In each step, the spinon can either hop to a neighbouring site or to stay at where it is. The number of hops is $L_\gamma$. We take the formal limit $\delta\tau\to 0$. Likewise, in the denominator, the summation is over all walks $\gamma$ that starts from $i$ at time 0 and arrives at any site at time $\tau$. $W_\gamma$ is the weighting factor.

We employ the Pauling approximation to estimate the weighting factor. Therefore,
\begin{align}
C_{ij}(\tau) \approx \frac{\sum_{\gamma:i\to j}W^\mathrm{Pauling}_\gamma(d\tau)^{L_\gamma}}{\sum_{\gamma:i}W^\mathrm{Pauling}_\gamma(d\tau)^{L_\gamma}}=\frac{e^{-3\tau}\sum_{\gamma:i\to j}W^\mathrm{Pauling}_\gamma(d\tau)^{L}}{e^{-3\tau}\sum_{\gamma:i}W^\mathrm{Pauling}_\gamma(d\tau)^{L}}\equiv \frac{A_{ij}(\tau)}{A_{i}(\tau)}.
\label{supp_pathint1}
\end{align}
$W^\mathrm{Pauling}_\gamma$ stands for the Pauling estimate of the weighting factor. Note $A_i(\tau) \equiv 1$ had we used the exact value of $W_\gamma$. However, since we use the Pauling estimate, $A_i(\tau)$ is no longer a constant.

We use continuous time Monte Carlo to sample both the numerator $A_{ij}(\tau)$ and the denominator $A_{i}(\tau)$ \cite{Svljuaasen2005}. To this end, we label the walk $\gamma$ as a sequence of time-ordered links traversed by the spinon, $i_1j_1,i_2j_2,\cdots$, and the hopping events happen at time $\tau_1,\tau_2,\cdots$. $s_{n}$ denotes the first $n$ steps of $\gamma$. We then consider the following continuous time random walk. The waiting time $\Delta$ between two hops is drawn from the exponential distribution
\begin{align}
r(\Delta)d\Delta = 3e^{-3\Delta}d\Delta\quad(\Delta>0).
\end{align}
The transition probability from $s_n$ to $s_{n+1}$ is given by
\begin{align}
T_{s_n\to s_{n+1}} =\frac{1}{N_{s_n}} \frac{W^\mathrm{Pauling}_{s_{n+1}}}{W^\mathrm{Pauling}_{s_n}}.
\end{align}
Here $W^\mathrm{Pauling}_{s_n}$ is the weighting factor of the first $n$ steps of the walk. In particular, $T_{s_n\to s_{n+1}}=0$ if $s_{n+1}\not\in\Gamma$. $N_{s_n}$ is the normalization factor to ensure $\sum_{s_{n+1}}T_{s_n\to s_{n+1}}=1$. The first step is drawn with probability $p_{s_1}=1/4$. 

We generate $\gamma$ by successively hopping the spinon according to the above random walk rules. We imagine the spinon carries a clock. We set the clock to time 0 initially and turn it forward after each hop. We stop the walk when the clock time exceeds $\tau$. 

The probability for the walk $\gamma$ is given by,
\begin{align}
p_{\gamma} & = p_{s_1}T_{s_1\to s_2}T_{s_2\to s_3}\cdots T_{s_{L-1} \to s_L}\cdot  r(\tau_1)d\tau \cdot r(\tau_2-\tau_1)d\tau \cdots r(\tau_{L}-\tau_{L-1})d\tau \int^\infty _{\tau-\tau_{L}} r(\Delta)d\Delta \nonumber \\
& = \frac{1}{4}\frac{W^\mathrm{Pauling}_{s_L}/W^\mathrm{Pauling}_{s_1}}{N_{s_1}N_{s_2}\cdots N_{s_{L-1}}} e^{-3\tau}(3d\tau)^L = \frac{3^{L-1}}{N_{s_1}N_{s_2}\cdots N_{s_{L-1}}} W^\mathrm{Pauling}_{s_L} e^{-3\tau}(d\tau)^L.
\label{supp_ctrw}
\end{align}
Note $\gamma = s_L$. Here $\tau_1, \tau_2-\tau_1, \cdots \tau_{L}-\tau_{L-1}$ are the waiting time between hops, and $r(\tau_1)d\tau, r(\tau_2-\tau_1)d\tau,\cdots r(\tau_L-\tau_{L-1})d\tau$ are their corresponding probabilities. Since the walk must stop at time $\tau$, the waiting time $\Delta$ between the $L$-th step and the $L+1$-th step must exceed $\tau-\tau_L$. The integral represents its probability. We have used $W^\mathrm{Pauling}_{s_1} = 3/4$ in the last equality.

It is important to note that the above random walk rules generate all $\gamma$ with non-zero weight. If $W^\mathrm{Pauling}_\gamma \neq 0$, then $W^\mathrm{Pauling}_{s_n}\neq 0$ for any sub-walk $s_n$. Hence $p_{\gamma}>0$. 

Comparing Eqs.~\eqref{supp_pathint1} and \eqref{supp_ctrw},
\begin{align}
A_{ij}(\tau) = \sum_{\gamma}\frac{N_{s_1}}{3}\frac{N_{s_2}}{3}\cdots \frac{N_{s_{L-1}}}{3} p_{\gamma}.
\label{supp_ctrw1}
\end{align}
We sample $A_{ij}(\tau)$ based on Eq.~\eqref{supp_ctrw1}. The denominator $A_{i}(\tau)$ is sampled in the same vein.

\section{Analytic continuation}\label{supp_AC}

We use analytic continuation to construct $C_\mathbf{q}(\omega)$ from the imaginary time data $C_\mathbf{q}(\tau)$,
\begin{align}
C_\mathbf{q}(\tau) = \int d\omega C_\mathbf{q}(\omega) e^{-\omega\tau}.
\label{supp_analytic}
\end{align}
$C_\mathbf{q}(\tau)$ is given by,
\begin{align}
C_\mathbf{q}(\tau) = \frac{A_\mathbf{q}(\tau)}{A_{\mathbf{q}=0}(\tau)},
\label{supp_analytic2}
\end{align}
where $A_{\mathbf{q}}(\tau)$ is the spatial Fourier transform of $A_{ij}(\tau)$.

The analytic continuation is performed by using the standard maximum entropy method \cite{Jarrell1996}. It is known that the maximum entropy method becomes less reliable at the high energy end of the spectrum. To remedy this, we use the first 7 power moments of $C_\mathbf{q}(\omega)$ to construct the default model \cite{Mead1984}. Note we use only the first moment to construct the default model for $\mathbf{q}=(\pi,0,0)$ because the spectrum is a simple delta peak at $(\pi,0,0)$.

The power moments of $C_\mathbf{q}(\omega)$ are related to the short-time expansion of $C_\mathbf{q}(\tau)$,
\begin{align}
C_{\mathbf{q}}(\tau) = c^{(0)}_{\mathbf{q}} - c^{(1)}_\mathbf{q}\tau+c^{(2)}_\mathbf{q}\frac{\tau^2}{2!} - c^{(3)}_\mathbf{q}\frac{\tau^3}{3!}+\cdots.
\end{align}
$c^{(k)}_\mathbf{q}$ is the $k$-th power moments of $C_{\mathbf{q}}(\omega)$. To find $c^{(k)}_\mathbf{q}$, we consider the short-time expansion of $A_\mathbf{q}(\tau)$, which is written as,
\begin{align}
A_\mathbf{q}(\tau) = e^{-3\tau}[a^{(0)}_\mathbf{q}+a^{(1)}_\mathbf{q}\tau+a^{(2)}_\mathbf{q}\frac{\tau^2}{2!}+a^{(3)}_\mathbf{q}\frac{\tau^3}{3!}+\cdots].
\end{align}
One can show that the expansion coefficients are given by,
\begin{align}
a^{(k)}_\mathbf{q} = \sum_{|s|=k}W^{\mathrm{Pauling}}_{s}\cdot \exp(i\mathbf{q}\cdot\mathbf{R}_s).
\label{supp_acoeff}
\end{align}
The summation is over all $k$-step walks $s$ on the diamond lattice. At each step, the spinon hops to a neighbouring site. Note that the spinon doesn't stay at the same site, which is slightly different from the path integral in the main text. $W^\mathrm{Pauling}_{s}$ is the weighting factor of the walk. $\mathbf{R}_s$ is the vector connecting the starting point and the end point of the walk $s$. We evaluate Eq.~\eqref{supp_acoeff} by brute-force enumeration. Once the coefficients $a^{(k)}_\mathbf{q}$ are known, we use Eq.\eqref{supp_analytic2} to calculate $c^{(k)}_\mathbf{q}$.

$C_\mathbf{q}(\tau)$ decays monotonically as $\tau$ increases. As we can see from Eq.~\eqref{supp_analytic}, $C_\mathbf{q}(\tau)$ decays faster if most of the spectral weight of $C_\mathbf{q}(\omega)$ is at the higher energy. Hence, along the high symmetry direction $\mathbf{q}=(q,0,0)$, it takes less efforts to obtain high precision Monte Carlo results for $q$ closer to $\pi$, where the spectrum is located near the top of the spinon energy band. We therefore use the Monte Carlo for $\pi/2<q<\pi$, and use the sublattice symmetry $C_\mathbf{q}(6-\omega) = C_{\mathbf{Q}-\mathbf{q}}(\omega)$ (see Section \ref{supm_matsym}) to obtain the spectrum for $0<q<\pi/2$. Here $\mathbf{Q}=(\pi,0,0)$. For $q=\pi/2$, we use the sublattice symmetry $C_\mathbf{Q/2}(\omega) = C_{\mathbf{Q}/2}(6-\omega)$ to constrain the maximum entropy analytic continuation process.

\section{Sublattice symmetry of $C_\mathbf{q}(\omega)$}\label{supm_matsym}

To see that $C_\mathbf{q}(\omega)$ does possess the sublattice symmetry, we consider the path integral representation of $C_\mathbf{q+Q}(\tau)$,
\begin{align}
C_{\mathbf{q}+\mathbf{Q}}(\tau) = e^{-3\tau}\sum_{\gamma} W_\gamma (d\tau)^L e^{i(\mathbf{q}+\mathbf{Q})\cdot\mathbf{R}_{ij}}.
\end{align}
$L$ is the number of spinon hops along the world line $\gamma$. On the one hand, we observe that the spinon alternatively occupies A and B sites. In our coordinates convention, the coordinates of the A sites are even integers, while the coordinates of the B sites are odd integers. Thus,
\begin{align}
C_{\mathbf{q}+\mathbf{Q}}(\tau) = e^{-3\tau} \sum_{\gamma} W_\gamma (d\tau)^L (-1)^L e^{i\mathbf{q}\cdot\mathbf{R}_{ij}} = e^{-3\tau} \sum_{\gamma} W_\gamma (-d\tau)^L e^{i\mathbf{q}\cdot\mathbf{R}_{ij}}
\label{supp_subsym1}
\end{align}
On the other hand, the path integral representation of $C_\mathbf{q}(-\tau)$ is given by,
\begin{align}
C_\mathbf{q}(-\tau) = e^{3\tau} \sum_{\gamma} W_\gamma (-d\tau)^L e^{i\mathbf{q}\cdot\mathbf{R}_{ij}}.
\label{supp_subsym2}
\end{align}
Comparing Eq.\eqref{supp_subsym1} and Eq.\eqref{supp_subsym2}, we find,
\begin{align}
C_{\mathbf{q}+\mathbf{Q}}(\tau) = e^{-6\tau} C_\mathbf{q}(-\tau).
\end{align}
Combining the above identity with Eq.~\eqref{supp_analytic}, we obtain
\begin{align}
C_{\mathbf{q}+\mathbf{Q}}(\omega) = C_\mathbf{q}(6-\omega).
\end{align}
Furthermore, the bond-centre inversion symmetry implies,
\begin{align}
C_{-\mathbf{q}}(\omega) = C_\mathbf{q}(\omega).
\end{align}
Combing the above two identities, we find,
\begin{align}
C_\mathbf{q}(6-\omega) = C_{\mathbf{Q}-\mathbf{q}}(\omega).
\end{align}

\section{Kernel polynomial method}

The Chebyshev polynomial expansion is another method to construct $C_\mathbf{q}(\omega)$ that is complementary to maximum entropy analytic continuation \cite{Weisse2006}. For our purpose, it is convenient to scale the frequency $\omega$ to the interval $[-1,1]$:
\begin{align}
w = \frac{\omega-3}{3}.
\end{align}
Here $w$ is the scaled frequency. In terms of the scaled frequency, the spectral function can be written as,
\begin{align}
\widetilde{C}_{ij}(w) = \mathcal{N}\langle\Psi| c^\dagger_j c_j \delta(w-H_c/3)c^\dagger_ic_i |\Psi\rangle.
\end{align}
Its $k$-th moment is given by,
\begin{align}
\int^1_{-1}w^k \widetilde{C}_{ij}(w)dw = \mathcal{N}\frac{\langle\Psi | c^\dagger_jc_j H^k_c c^\dagger_i c_i |\Psi\rangle}{3^k} = (-1)^k \frac{\langle\Psi | c^\dagger_j c_j H^k_c c^\dagger_i c_ i |\Psi\rangle}{\langle\Psi|H^k_c c^\dagger_i c_i|\Psi\rangle}.
\end{align}
In the last equality, we have used the fact that $H_c|\Phi\rangle = -3|\Phi\rangle$ and $\langle\Psi| c^\dagger_ic_i |\Psi \rangle = 1/\mathcal{N}$. Using the same idea in Appendix \ref{supp_AC}, the moments can be expressed as a summation over spinon walks,
\begin{align}
\int^1_{-1} w^k \widetilde{C}_\mathbf{q}(w)dw = (-1)^k\frac{\sum_{|s|=k} W_s \exp(i\mathbf{q}\cdot\mathbf{R}_{s})}{\sum_{|s|=k} W_s} \approx (-1)^k \frac{\sum_{|s|=k} W^\mathrm{Pauling}_s \exp(i\mathbf{q}\cdot\mathbf{R}_{s})}{\sum_{|s|=k} W^\mathrm{Pauling}_s}.
\end{align}
The summation is over all walks of $k$ steps. At each step, the spinon hops to a neighbouring site. $\mathbf{R}_{s}$ is the vector connecting the starting point and the end point of the walk $s$. We evaluate the first 16 power moments of $\widetilde{C}_\mathbf{q}(w)$ by enumerating all possible walks up to 15 steps. We then construct $\widetilde{C}_\mathbf{q}(w)$ from its moments by using the Chebyshev polynomial expansion. Finally, we find $C_\mathbf{q}(\omega)$ through the relation $ C_\mathbf{q}(\omega) = 3\widetilde{C}_\mathbf{q}[(\omega-3)/3]$.

\end{document}